\documentclass[journal]{IEEEtran}
\ifCLASSINFOpdf
\else
\fi
\usepackage{amsmath}
\usepackage{url}

\usepackage{amsfonts, amssymb}
\usepackage{graphicx}
\usepackage{booktabs}
\usepackage{xcolor}
\usepackage{algpseudocode}

\begin{document}
%
\title{A Study of Parallelizable Alternatives to Dynamic Time Warping for Aligning Long Sequences}
%
%
%

\author{Daniel~Yang,$^*$
        Thaxter~Shaw,$^*$
        and~TJ~Tsai,~\IEEEmembership{Member,~IEEE}
\thanks{$^*$ Equal contribution.  D. Yang and T. Shaw are students at Harvey Mudd College.}
\thanks{T. Tsai is with the Department of Engineering, Harvey Mudd College, Claremont,
CA, 91711 USA e-mail: ttsai@g.hmc.edu.}
\thanks{\copyright~2022 IEEE. Personal use of this material is permitted. Permission from IEEE must be obtained for all other uses, in any current or future media, including reprinting/republishing this material for advertising or promotional purposes, creating new collective works, for resale or redistribution to servers or lists, or reuse of any copyrighted component of this work in other works.}%
}

\maketitle

\begin{abstract}
This article investigates several parallelizable alternatives to DTW for estimating the alignment between two long sequences.  Whereas most previous work has focused on reducing the total computation and/or memory costs of DTW, our focus is instead on reducing wall clock time by utilizing common hardware like GPUs that are optimized for parallel processing.  We propose and study four different parallelizable alignment algorithms: the first three algorithms compute approximations of DTW by breaking the pairwise cost matrix into rectangular regions and processing the regions in parallel, and the fourth algorithm computes an exact DTW alignment by processing the cost matrix along diagonals rather than rows or columns.  We characterize the performance of our proposed alignment algorithms on an audio-audio alignment task, and we develop GPU-based implementations for the two best-performing algorithms, which we call weakly-ordered Segmental DTW (WSDTW) and Parallelized Diagonal DTW (ParDTW).  Our experiments indicate that ParDTW is the most practical and useful of the four algorithms: it computes an exact DTW alignment and reduces runtime by $1.5$ to $2$ orders of magnitude on long sequences compared to current alternatives.  We present a comprehensive evaluation and study of the alignment accuracy, runtime, and practical limitations of the proposed alignment algorithms.
\end{abstract}


%
\IEEEpeerreviewmaketitle

\section{Introduction}
%
%
%
%

\IEEEPARstart{T}{his} paper explores parallelizable alternatives to dynamic time warping (DTW). DTW is a dynamic programming algorithm for determining the optimal alignment between two sequences of features.  It was originally developed in the context of speech recognition \cite{sakoe1978dynamic} and is now widely used in many tasks involving time series data \cite{dau2019ucr}.  Its weaknesses are its quadratic runtime and memory costs, which can become prohibitively expensive for long sequences.  In this paper, we explore parallelizable alternatives to DTW that make use of modern hardware such as graphics processing units (GPUs).

\textcolor{black}{Recent work in the machine learning community has explored ways to incorporate DTW alignment into neural network models with sequence inputs.  Some works integrate DTW into a neural network by estimating an alignment, and then only backpropagating through the optimal alignment path in the pairwise cost matrix \cite{cai2019dtwnet}\cite{iwana2020dtw}.  This approach has also been explored in the past with other feature transformations (e.g.~CCA \cite{zhou2009canonical}).  Other works approach the problem by considering smooth, differentiable approximations of DTW, and then backpropagating through the estimated soft alignments \cite{cuturi2017softdtw}\cite{mensch2018differentiable}\cite{blondel2021differentiable}.  The use of soft alignments has been successfully applied in several neural network models (e.g.~\cite{chang2019d3tw}\cite{cao2020few}), but the use cases have been limited to very short sequences (e.g.~$8 \times 8$ pairwise cost matrix in \cite{cao2020few}) due to the computational cost of DTW.  Given that most modern neural networks are trained on parallelized hardware like GPUs and require very fast per-batch processing times, this motivates our exploration of parallelizable alternatives to DTW that can utilize modern GPUs and be fast enough to be incorporated into model training.}

Many previous works have proposed ways to make DTW more scalable.  These works can be divided into two categories.  The first category propose ways to speed up an exact DTW alignment.  Many works focus on efficient computation of the DTW 1-nearest neighbor for short scalar sequences through the use of lower bounds \cite{keogh2005exact}\cite{ding2008querying}\cite{keogh2009supporting}\cite{zhou2011boundary}, early abandoning \cite{rakthanmanon2012searching}\cite{li2007ea}\cite{li2019speed}, and utilizing multiple cores \cite{shabib2015parallelization}\cite{movchan2015parallel}\cite{atia2019hand} or specialized hardware \cite{wang2013accelerating}\cite{sart2010accelerating}\cite{zhang2012fast}.  Much of this work revolves around the UCR Time Series Archive \cite{dau2019ucr}, which focuses on classification tasks of short scalar sequences (ranging in length from 15 to 2844).  Lower bounds for multi-dimensional time series do exist \cite{yang2013tighter}\cite{zhang2011inner}\cite{shen2021tc}\cite{li2017distance}, but are more limited in scope.  For alignment of long sequences, a recent work \cite{gold2018dynamic} proposed a two-stage method for computing DTW under certain conditions in $O(L^2 log(log(log(L)))/log(log(L))$ time.  Another recent work \cite{tralie2020exact} proposes a way to reduce memory costs to $O(L)$ by performing dynamic programming along diagonals (rather than rows/columns) of the cost matrix, combined with a divide-and-conquer approach.

The second category of previous work proposes approximations to DTW that require less computation or memory.  For alignment of short query sequences, many works have explored approximate lower bounds \cite{tavenard2015improving}\cite{lee2005approximate}\cite{zhang2011piecewise} or approximations of DTW distance \cite{lods2017learning}\cite{tan2017indexing}\cite{nagendar2015efficient}.  For alignment of long sequences, some methods include imposing bands in the pairwise cost matrix to limit extreme time warping \cite{dau2018optimizing}\cite{sakoe1978dynamic}\cite{itakura1975minimum}, adopting a multi-resolution approach \cite{salvador2007toward}\cite{MuellerMK06_EfficientMultiscaleApproach_ISMIR} in which an alignment is estimated at a coarse granularity and then iteratively refined, and estimating alignments given a fixed amount of memory \cite{PraetzlichDM16_MsDTW_ICASSP} or in an online fashion \cite{macrae2010accurate}\cite{dixon2005live}.

This paper explores parallelizable alignment algorithms for aligning long sequences.  Whereas previous works focus on reducing total computation and memory costs, our focus is on reducing wall clock time (rather than total computation) by utilizing common hardware like GPUs that are optimized for parallel processing.  We propose and investigate four parallelizable alignment algorithms.  The first three are variants of an algorithm called Segmental DTW \cite{tsai2017make} that compute different approximations of DTW.  Segmental DTW was originally proposed to solve a completely different problem -- estimating the alignment between a long reference recording and an ordered set of audio fragments separated by unknown gaps -- and we adapt its original formulation to approximate DTW in a parallelizable way.  The 3 variants we investigate differ in the ordering constraints imposed on the estimated alignment path, and are appropriately named non-ordered Segmental DTW (NSDTW), weakly-ordered Segmental DTW (WSDTW), and strictly-ordered Segmental DTW (SSDTW).  In all 3 variants, the global alignment problem is broken into smaller alignment problems that can be solved in parallel.  The fourth alignment algorithm is called Parallelized Diagonal DTW (ParDTW) and computes an exact DTW alignment in a parallelizable way.  It uses a recent insight \cite{tralie2020exact} that performing dynamic programming along diagonals of the cost matrix (rather than along rows or columns) allows for parallel computations, but it focuses on minimizing runtime rather than minimizing memory usage (as in \cite{tralie2020exact}).

This article has three main contributions.\footnote{\textcolor{black}{This article is a journal extension to an earlier conference paper \cite{tsai2021segmental}.  It extends the conference paper in several ways: (i) we propose two additional parallelizable alignment algorithms (NSDTW in Section \ref{subsec:nsdtw}, ParDTW in Section \ref{subsec:pardtw}) and characterize the performance of all proposed algorithms and additional baselines in both alignment accuracy (Figure \ref{fig:results}) and runtime (Section \ref{subsec:analysis_runtime_cpu}), (ii) we propose a highly parallelized implementation of WSDTW (Section \ref{subsec:gpuImpl_wsdtw}) that parallelizes along two additional dimensions beyond the conference paper, (iii) we release the source code for highly optimized GPU-based implementations of WSDTW and ParDTW, whereas the conference paper only had optimized CPU-based implementations, and (iv) we conduct several new analyses including the effect of SNR (Section \ref{subsec:analysis_snr}), wall clock runtime of parallelized implementations (Section \ref{subsec:analysis_runtime_gpu}), and memory requirements (Section \ref{subsec:analysis_memory}).}} First, we propose four parallelizable alignment algorithms for aligning long sequences that could be used as alternatives to DTW: non-ordered Segmental DTW, weakly-ordered Segmental DTW, strictly-ordered Segmental DTW, and Parallelized Diagonal DTW.  Second, we characterize the performance of these alignment algorithms on an audio-audio alignment task and under various controlled conditions.  Our empirical results indicate that Parallelized Diagonal DTW is the most practical and useful of the four algorithms: it computes an exact DTW alignment and reduces runtime by $1.5$ to $2$ orders of magnitude on long sequences compared to current alternatives.  Third, we release an open source implementation of Parallelized Diagonal DTW that is optimized for use with GPUs.  Code can be found at \url{https://github.com/HMC-MIR/ParallelizingDTW}.

The rest of the paper is organized as follows.  Section \ref{sec:systemDescr} describes the four proposed parallelizable alignment algorithms.  Section \ref{sec:gpuImpl} provides further detail on how these algorithms can be implemented efficiently on a GPU.  Section \ref{sec:expSetup} explains the experimental setup for an audio-audio alignment task, and Section \ref{sec:results} presents our empirical results.  Section \ref{sec:analyses} conducts several analyses to provide deeper intuition into algorithmic behavior.  Section \ref{sec:concl} concludes the work.

\section{System Description}
\label{sec:systemDescr}

This section describes four different alignment algorithms that could be used as parallelizable alternatives to DTW.  The first three alignment algorithms are variants of Segmental DTW and are all approximations of DTW.  The fourth alignment algorithm is a parallelized version of DTW which computes an exact DTW alignment.  These four alignment algorithms will be described in detail in the next four subsections.

\begin{figure}[!t]
	\centering
	\includegraphics[width=\columnwidth]{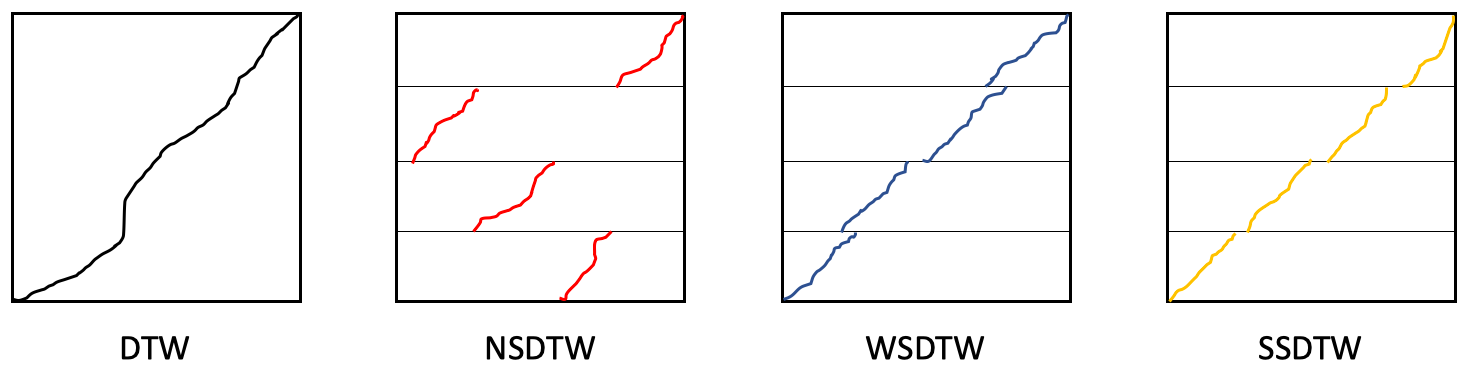}
	\caption{Sample alignment paths for DTW, NSDTW, WSDTW, and SSDTW.  NSDTW imposes no ordering constraints on the subsequence alignment paths and may have large forward or backward jumps at fragment boundaries.  WSDTW imposes weak ordering constraints, allowing forward jumps or short backward jumps at fragment boundaries.  SSDTW imposes strict ordering constraints, allowing forward jumps only at fragment boundaries.}
	\label{fig:sample_alignments}
\end{figure}

\subsection{Non-ordered Segmental DTW}
\label{subsec:nsdtw}
The first alignment algorithm is non-ordered Segmental DTW (NSDTW).  Aligning two sequences A and B with NSDTW consists of two steps.  The first step is to break sequence A into $N$ approximately equal-length fragments.  The second step is to perform subsequence DTW between each fragment of sequence A and the entirety of sequence B.  Subsequence DTW is a variant of DTW that finds the optimal alignment between a short query sequence and any subsequence within a long reference sequence.  It accomplishes this by (1) computing a pairwise cost matrix between the query sequence and reference sequence, (2) initializing the cumulative cost matrix to allow the alignment path to begin anywhere in the reference sequence without penalty, (3) filling out the cumulative cost matrix and corresponding backtrace matrix using dynamic programming, and (4) using the lowest subsequence path score as the starting point for backtracing in order to determine the optimal subsequence alignment path.  \textcolor{black}{For a more detailed explanation of subsequence DTW, the reader is referred to \cite{muller2007information}, section 4.4.}  In our experiments, the dynamic programming stage allows for (query, reference) transitions $\{(1, 1), (1, 2), (2, 1)\}$ with corresponding multiplicative weights $\{1, 1, 2\}$.  Note that the subsequence DTW computations for each fragment can be done in parallel.  The final predicted alignment from NSDTW is simply the concatenation of the optimal subsequence paths for each fragment.

NSDTW has a major weakness: it imposes no ordering constraints on the subsequence alignment paths.  This can result in sudden discontinuities at the fragment boundaries, as shown in Figure \ref{fig:sample_alignments}.  \textcolor{black}{For example, if the first fragment in sequence A has several strong matches in sequence B, NSDTW will simply select the strongest match without any consideration of global ordering among the fragment alignments.}  Whereas regular DTW produces a monotonic alignment path, NSDTW does not guarantee continuity or monotonicity and may have large forward or backward jumps.  This potential weakness will be addressed by the weakly-ordered and strictly-ordered variants described in the next two subsections.

\begin{figure}[!t]
	\centering
	\includegraphics[width=\columnwidth]{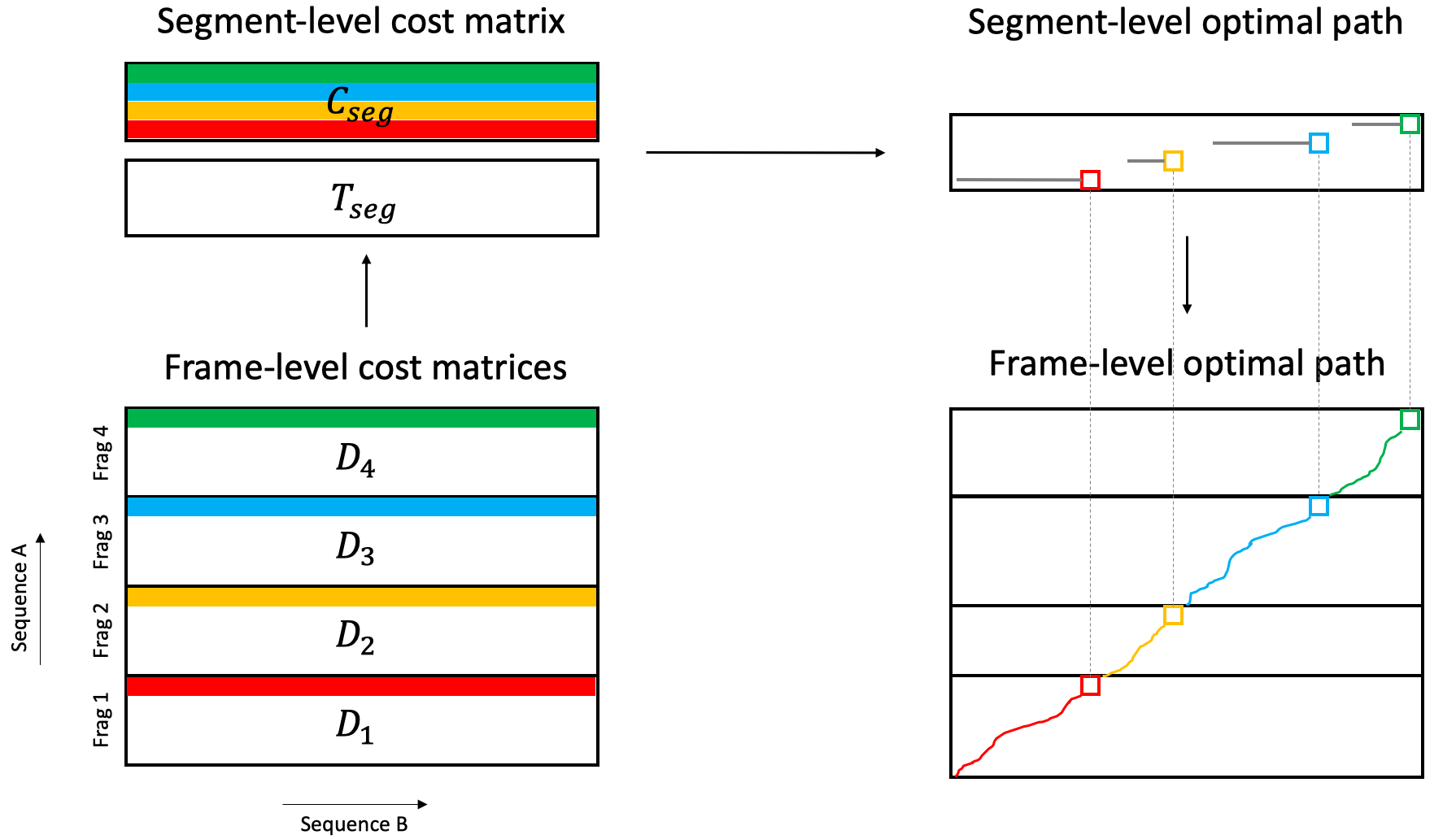}
	\caption{Overview of the main steps in weakly-ordered and strictly-ordered Segmental DTW.}
	\label{fig:wsdtw_overview}
\end{figure}

\subsection{Weakly-ordered Segmental DTW}
\label{subsec:wsdtw}

The second alignment algorithm is weakly-ordered Segmental DTW (WSDTW).  It partially addresses the weakness of NSDTW by imposing some ordering constraints.  Aligning two sequences A and B with WSDTW consists of five steps, which are described in the next five paragraphs.  Figure \ref{fig:wsdtw_overview} shows an overview of the algorithm, where the sequences being aligned are assumed to go from bottom to top and from left to right.

The first step is to break sequence A into $N$ approximately equal-length fragments.  This step is identical to the first step of NSDTW.  Let $L_A$ and $L_B$ denote the lengths of sequences A and B, respectively, and consider the pairwise cost matrix $C_i$ between the $i^{th}$ fragment of sequence A and the entirety of sequence B.  The matrices $C_1, C_2, \dots, C_N \in \mathbb{R}^{L_A/N \times L_B}$ are simply a partition of the global pairwise cost matrix $C \in \mathbb{R}^{L_A \times L_B}$ into $N$ sub-matrices.  Figure \ref{fig:wsdtw_overview} shows an example of this step with $N=4$ fragments.

The second step is to compute a subsequence DTW cumulative cost matrix on each fragment.  For each pairwise cost matrix $C_i \in \mathbb{R}^{L_A/N \times L_B}$, $i=1,2,\dots,N$, we compute a corresponding cumulative cost matrix $D_i \in \mathbb{R}^{L_A/N \times L_B}$ that indicates subsequence path scores, as well as the corresponding backtrace matrix $B_i \in \mathbb{N}^{L_A/N \times L_B}$.  As before, we allow (query, reference) transitions $\{(1, 1), (1, 2), (2, 1)\}$ with corresponding multiplicative weights $\{1, 1, 2\}$, where the fragment of sequence A serves as the query and sequence B serves as the reference.  Note that this second step can be done in parallel across the $N$ fragments.  This step is identical to the second step of NSDTW, except that instead of selecting locally optimal subsequence alignment paths, we will perform some additional computations (steps 3 and 4, described below) to select a set of globally optimal subsequence alignment paths under certain constraints.

The third step is to assemble a segment-level cost matrix $C_{seg}$. This is done by stacking the subsequence path scores from each $D_i$, $i=1,2,\dots,N$ into a matrix of size $\mathbb{R}^{N \times L_B}$.  The subsequence path scores are shown in Figure \ref{fig:wsdtw_overview} as highlighted rows at the top of each sub-matrix $D_i$.  $C_{seg}[i,j]$ therefore indicates the optimal subsequence path score of the $i^{th}$ fragment ending at offset $j$ in sequence B. $C_{seg}$ will serve as the segment-level pairwise cost matrix.

The fourth step is to find an optimal path through $C_{seg}$ that enforces ordering constraints.  This can be accomplished with dynamic programming with allowable (fragment, frame) transitions $\{(0, 1), (1, \frac{L_A}{2N})\}$ and corresponding multiplicative weights $\{0, 1\}$.  In Figure \ref{fig:wsdtw_overview}, the $(0,1)$ transition corresponds to a horizontal transition directly to the right, which indicates a skip with no cost.  The $(1, \frac{L_A}{2N})$ transition corresponds to a segment-level match, where $\frac{L_A}{2N}$ represents the smallest possible matching duration with a fragment of length $\frac{L_A}{N}$ and a maximum time warping factor of 2 (since the allowable frame-level transitions are $(1,1)$, $(1,2)$ and $(2,1)$).  The dynamic programming equation is thus given by $D_{seg}[i,j] = min(D_{seg}[i,j-1], D_{seg}[i-1, j - \frac{L_A}{2N}] + C_{seg}[i,j])$.  Once the segment-level cumulative cost matrix $D_{seg}$ and corresponding backtrace matrix $B_{seg}$ have been computed, we can use the backpointers in $B_{seg}$ to determine the optimal path through $C_{seg}$.  This segment-level path indicates the ending locations for a set of globally optimal subsequence alignment paths.  In Figure \ref{fig:wsdtw_overview}, these ending locations are indicated with colored empty boxes.  Note that these ending locations are guaranteed to be strictly increasing, which ensures that the fragments are globally ordered.

The fifth step is to backtrace through each frame-level backtrace matrix $B_i$.  The backtrace for each fragment starts at the globally optimal ending location for the given fragment, as determined in the previous step.  This backtracing step can be done in parallel across each fragment.  In Figure \ref{fig:wsdtw_overview}, the ending locations are shown as colored empty boxes, and the backtraced alignment paths are indicated with colored lines.  The concatenation of subsequence alignment paths is the final predicted alignment of the WSDTW algorithm.  

Note that WSDTW only imposes weak ordering constraints.  Whereas regular DTW with $\{(1,1)$, $(1,2)$, $(2,1)\}$ transitions guarantees a smooth, monotonic alignment path, WSDTW allows for discontinuities at the fragment boundaries, including the possibility of backward jumps.  To see this, consider the most extreme example: the ending locations of two consecutive fragment alignment paths are separated by the minimum distance $\frac{L_A}{2N}$, but the backtraced frame-level alignment path for the later fragment consists entirely of $(1,2)$ transitions and thus has a duration (along sequence B) of $\frac{2L_A}{N}$.  In this case, there will be a backward jump of magnitude $\frac{2L_A}{N} - \frac{L_A}{2N} = \frac{3L_A}{2N}$ at the boundary between the two fragments.  While this example would probably never occur in practice, we do find that it is not uncommon to have small backward jumps at fragment boundaries.

It is useful to point out that the frame-level operations (steps 1, 2, and 5) in WSDTW are parallelizable.  In section \ref{sec:gpuImpl}, we will describe how WSDTW can be implemented efficiently on a GPU.

\begin{figure}
	\begin{algorithmic}
		\small
		\Procedure{SDTW}{SeqA, SeqB, $L_A$, $L_B$, $N$, steps, wts, variant}
		\For{i from 0 to $N$}
		\State frag $\gets$ \Call{getIthFragment}{SeqA, $N$, i}
		\State D[i], B[i] $\gets$ \Call{subseqDTW}{frag, SeqB, steps, wts}
		\EndFor
		\vspace{6pt}
		\State $C_{seg}$, $T_{seg}$ $\gets$ \Call{getSegLevelMatrices}{D, B, variant}
		\State fragEnds $\gets$ \Call{getFragEndLocs}{$C_{seg}$, $T_{seg}$, variant}
		\vspace{6pt}
		\For{i from 0 to $N$}
		\State fragAlign $\gets$ \Call{backtraceFrom}{B[i], fragEnds[i]}
		\State fragAlign $\gets$ \Call{addOffset}{fragAlign, i $\times$ $L_A/N$}
		\State path $\gets$ \Call{extend}{path, fragAlign}
		\EndFor
		\State \textbf{return} path
		\EndProcedure
	\end{algorithmic}
	\caption{\textcolor{black}{Pseudo code for Segmental DTW.  Both for loops can be parallelized. 
		The \textsc{getSegLevelMatrices} and \textsc{getFragEndLocs} functions behave differently for NSDTW, WSDTW, and SSDTW to enforce different global ordering constraints.
		NSDTW chooses the minimum element in the top row of each $D_i$.
		WSDTW and SSDTW use dynamic programming to find globally optimal fragment ending locations.
		SSDTW uses $T_{seg}$ to define the allowable transitions through $C_{seg}$.}}
	\label{fig:segmentaldtw_pseudocode}
\end{figure}

\subsection{Strictly-ordered Segmental DTW}
\label{subsec:ssdtw}

The third alignment algorithm is strictly-ordered Segmental DTW (SSDTW).  It addresses the weakness of NSDTW by imposing strict ordering constraints.  Unlike WSDTW, SSDTW guarantees that the global alignment path will not have backward jumps.  SSDTW consists of the same five steps as shown in Figure \ref{fig:wsdtw_overview}, but it has several differences from WSDTW which are described below.

The first two steps are identical to WSDTW: sequence A is broken into $N$ fragments, and a subsequence DTW cumulative cost matrix is calculated on each fragment.  These two steps can be done in parallel across each fragment.

The third step is to assemble the segment-level cost matrix $C_{seg}$ and a segment-level transition matrix $T_{seg}$.  $C_{seg}$ is formed in the same way as in WSDTW.  The difference in this third step is that SSDTW requires constructing an additional matrix $T_{seg} \in \mathbb{N}^{N \times L_B}$ that keeps track of valid transitions between elements in $C_{seg}$.  $T_{seg}[i, j]$ indicates the starting position (along sequence B) of the optimal alignment path for the $i^{th}$ fragment that ends at position $j$ in sequence B.  For example, if $k \triangleq T_{seg}[i,j]$, then there is a valid transition directly from position $(i-1, k)$ to position $(i,j)$ in $C_{seg}$.  Computing all entries of $T_{seg}$ therefore requires backtracing from every possible ending location for every fragment.

The fourth step is to find the optimal path through $C_{seg}$ using valid transitions in $T_{seg}$.  This can be accomplished with dynamic programming with two types of allowable transitions.   The first type is a $(0,1)$ transition with multiplicative weight 0 that indicates a horizontal skip with no cost.  The second type is a jump transition with multiplicative weight 1 that is determined by the information in $T_{seg}$.  If we define $k \triangleq T_{seg}[i,j]$, then the dynamic programming equation is given by $D_{seg}[i,j] = min(D_{seg}[i,j-1], D_{seg}[i-1,k] + C_{seg}[i,j])$.  Note that, whereas WSDTW allows a fixed $(1, \frac{L_A}{2N})$ jump transition at each position, SSDTW allows a jump transition that is specific to each position $(i,j)$ in $C_{seg}$ in order to ensure that backward jumps cannot occur.

The fifth step is to backtrace through each frame-level backtrace matrix $B_i$.  This step is identical to WSDTW: the backtrace for each fragment starts at the globally optimal ending location for the given fragment, and backtracing is used to determine the optimal subsequence alignment path.  The concatenation of each fragment's subsequence alignment path forms the final global predicted alignment.  This step can be done in parallel across each fragment.

Note that SSDTW guarantees a monotonic global alignment path.  It still allows forward jumps at fragment boundaries, but the transition matrix guarantees that each fragment's alignment path does not begin before the previous fragment's alignment path has ended.  This guarantee comes at the cost of significant additional computation in calculating the entries of $T_{seg}$.

\textcolor{black}{Figure \ref{fig:segmentaldtw_pseudocode} shows pseudo code for all three Segmental DTW variants.  Note that both for loops can be parallelized, and that the manner in which the fragment ending locations are selected varies based on which variant of Segmental DTW is being used.}

\subsection{Parallelized Diagonal DTW}
\label{subsec:pardtw}

The fourth alignment algorithm is Parallelized Diagonal DTW (ParDTW).  Unlike the previous three alignment algorithms, ParDTW is an exact implementation of DTW, but it performs the dynamic programming in a way that allows for parallelization on a GPU.  As a secondary objective, it is also designed to minimize memory usage, since GPUs have limited RAM.

ParDTW is identical to standard DTW but has three significant differences.  First, the dynamic programming is performed along diagonals of the cost matrix, rather than along rows and columns as in standard DTW.  As can be seen in the bottom illustration in Figure \ref{fig:parallelize_dims}, cumulative cost elements on each diagonal can be computed in parallel.  Second, the pairwise cost matrix is never allocated in memory.  Instead, each pairwise cost is computed on the fly when its value is needed.  By performing the pairwise cost computations on the fly rather than as a preprocessing step, we can eliminate the need to allocate the entire global pairwise cost matrix in memory.  Third, the cumulative cost matrix is never allocated in memory.  Instead, four fixed-length buffers are used to keep track of the cumulative cost values in the four most recent diagonals.\footnote{The number of buffers required depends on the set of allowable transitions.  In our description of ParDTW, we assume a set of transitions $\{(1,1), (1,2), (2,1)\}$.}  As shown in Figure \ref{fig:parallelize_dims} (bottom), calculating the cumulative cost values on a diagonal only requires knowing the values in the three previous diagonals.  By cyclically using four buffers, we can eliminate the need to allocate the entire global cumulative cost matrix in memory.  Apart from the three differences above, ParDTW is identical to standard DTW.  In particular, it is important to note that the backtrace matrix is still allocated in memory and used to perform backtracing.

The concept of using diagonals to parallelize DTW is not new.  Tralie and Dempsey \cite{tralie2020exact} introduced this idea in an algorithm that computes an exact DTW alignment but reduces the memory requirement from quadratic to linear.  To accomplish this, they use diagonal buffers with a divide-and-conquer approach that finds the midpoint of the DTW alignment path, and then applies the algorithm recursively to each of the two resulting (smaller) alignment problems until the entire alignment path has been determined.  ParDTW uses the insight that diagonal buffers can be parallelized, but it prioritizes minimizing total runtime (through a parallelized GPU implementation) rather than minimizing total memory usage.  As will be seen in Section \ref{sec:analyses}, it achieves a significant speedup in runtime but retains a quadratic memory requirement.

\section{GPU Implementations}
\label{sec:gpuImpl}

In this section we describe optimized GPU-based implementations of WSDTW and ParDTW.  As will be seen in Section \ref{sec:results}, WSDTW and ParDTW are the alignment algorithms with the best alignment accuracy.  We therefore focused on developing optimized GPU implementations of these two algorithms, the details of which are described in the next two subsections.

\subsection{GPU Implementation of WSDTW}
\label{subsec:gpuImpl_wsdtw}

The algorithm described in this subsection exactly implements the WSDTW algorithm described in Section \ref{subsec:wsdtw}, but it does so in a way that utilizes massive parallelization.  Our goal is to reduce total runtime by utilizing as much parallelization as possible on a GPU.

\begin{figure}[!t]
	\centering
	\includegraphics[width=\columnwidth]{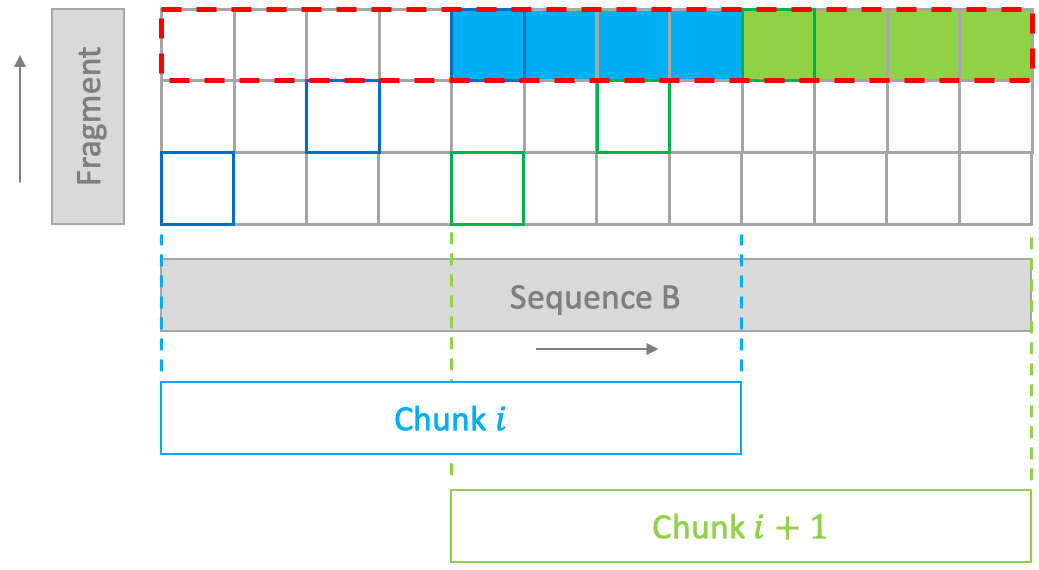}
	\caption{An illustration of how to parallelize subsequence DTW between a short fragment and a long reference sequence.  To compute the subsequence path scores colored in blue, only the blue chunk (``Chunk $i$") of the reference sequence is needed.  The computation can thus be parallelized across chunks.}
	\label{fig:parallelize_chunks}
\end{figure}

Recall from Figure \ref{fig:wsdtw_overview} that WSDTW consists of five stages: (1) breaking sequence A into $N$ fragments, (2) computing a subsequence DTW cumulative cost matrix between each fragment and the entirety of sequence B, (3) assembling a segment-level cost matrix $C_{seg}$, (4) finding an optimal path through $C_{seg}$, and (5) backtracing through each frame-level backtrace matrix $B_i$.  The segment-level operations (steps 3 and 4) are not parallelizable, and are simply run on a single core on the GPU.  The frame-level backtracing (step 5) is parallelized across the $N$ fragments in a straightforward manner.  Unlike steps 3 through 5, however, the frame-level dynamic programming (steps 1 and 2) can be massively parallelized.  The remainder of this section describes how this parallelization can be accomplished.

The frame-level dynamic programming is parallelized along three dimensions, which are described in the following three paragraphs.

The first dimension is to parallelize across the $N$ fragments in sequence A, as has already been discussed in Section \ref{subsec:wsdtw}.  For each fragment, we must compute a subsequence DTW between the fragment and the entirety of sequence B.

The second dimension is to parallelize across $M$ overlapping chunks in sequence B.  Note that when the allowable transitions are $\{(1,1), (1,2), (2,1)\}$, it is possible to compute a subsequence path score by looking at only a small context within sequence B.  Figure \ref{fig:parallelize_chunks} shows an illustration of this for a toy example, where we are calculating subsequence path scores (top row of the cumulative cost matrix, indicated with red dotted line) between a short fragment and the entirety of sequence B.  If our goal is to compute the individual subsequence path scores colored in blue, we only need to consider the context in B indicated by the blue chunk (``Chunk $i$").  Note that the potential path with most extreme time warping is indicated with blue highlighted boxes.  Similarly, if our goal is to compute the individual subsequence path scores colored in green, we only need to consider the context in B indicated by the green chunk (``Chunk $i+1$").  Therefore, we can parallelize each global subsequence DTW problem (i.e.~between a fragment and the entirety of sequence B) into a set of $M$ local subsequence DTW problems (i.e.~between a fragment and a chunk of sequence B).  Because the chunks are overlapping, this adds some additional redundant computation but allows us to potentially reduce runtime through parallelization.  Note that this process yields an exact computation -- it is not an approximation but yields the exact same results as computing a single global subsequence DTW.

\begin{figure}[!t]
	\centering
	\includegraphics[width=\columnwidth]{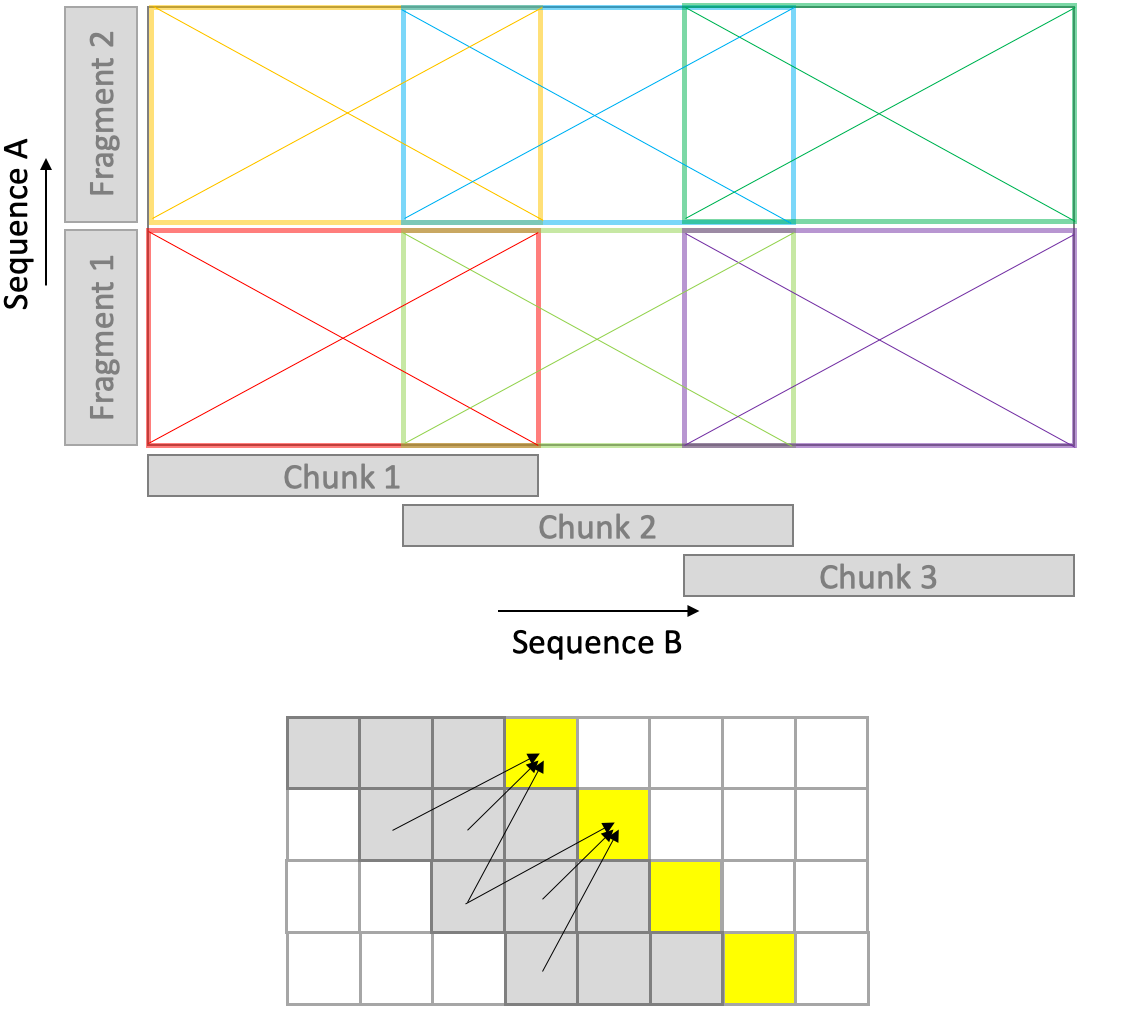}
	\caption{The dimensions across which WSDTW parallelizes computation.  (Top) The global pairwise cost matrix is divided into overlapping tiles, and each tile is processed in parallel.  (Bottom) Within each tile, subsequence DTW is performed by processing diagonal elements in parallel.}
	\label{fig:parallelize_dims}
\end{figure}

The third dimension is to parallelize each local subsequence DTW problem across diagonals.  As mentioned in Section \ref{subsec:pardtw}, this idea was recently proposed by Tralie and Dempsey \cite{tralie2020exact}, who note that by performing the dynamic programming across diagonals of a cost matrix (rather than rows or columns), it is possible to compute the elements on each diagonal in parallel.  Furthermore, we also adopt their technique of reducing memory storage by only keeping a few diagonal buffers in memory during the dynamic programming, rather than storing the entire cumulative cost matrix.\footnote{In their paper, Tralie and Dempsey \cite{tralie2020exact} assume $\{(1,0), (0,1), (1,1)\}$ transitions and only need three diagonal buffers.  In our paper, we assume $\{(1,1), (1,2), (2,1)\}$ transitions and need four diagonal buffers.}  Therefore, a (frame-level) pairwise cost matrix and a cumulative cost matrix are never allocated in memory during the frame-level dynamic programming.  Instead, pairwise costs are computed on the fly and only four buffers are used to keep track of the cumulative cost elements on the four most recent diagonals.  However, the entire global frame-level backtrace matrix must be stored in memory, since that information is needed to perform the backtracing in step 5.

Figure \ref{fig:parallelize_dims} illustrates the three dimensions along which we parallelize the frame-level dynamic programming.  One can think of the process as parallelizing across $NM$ overlapping tiles in the global pairwise cost matrix, and then computing the subsequence DTW on each tile by processing diagonal elements in parallel.  Note that the amount of parallelization is $NM$ times the maximum diagonal length in each tile $\frac{L_A}{N}$, which results in a parallelization of $L_A M$.  The combination of these three dimensions allows for massive parallelization on long sequences.

\subsection{GPU Implementation of ParDTW}
\label{subsec:gpuImpl_pardtw}

The GPU implementation of ParDTW is far simpler than WSDTW: it simply parallelizes across the diagonals of the cost matrix during the dynamic programming stage.  Its implementation is exactly the same as in WSDTW (paragraph 6 in the previous subsection), except that it computes a standard DTW alignment rather than a subsequence DTW alignment.

\section{Experimental Setup}
\label{sec:expSetup}

In this section, we describe the data and evaluation methodology for our experiments.

We ran all experiments on the Chopin Mazurka dataset \cite{sapp2008hybrid}. This dataset contains multiple audio performances of 5 different Chopin mazurkas, along with beat-level timestamp annotations for each performance.  Because DTW, Segmental DTW, and ParDTW do not have any trainable parameters, only one mazurka (Opus 17, No 4) was reserved as a training set for development and debugging, and the other 4 were used as a testing set.  Pairwise cost matrices were computed using cosine distance on standard chroma features with 23ms hop size.  In our experiments, we align each performance to all other performances of the same mazurka, resulting in a total of 1953 training pairs and 7630 testing pairs. Table \ref{tab:data} shows an overview of the dataset.

\begin{table}
	\caption{Overview of the Chopin Mazurka dataset used in alignment experiments.  Durations are indicated in seconds.}
	\label{tab:data}
	\begin{center}
		\begin{tabular}{|l|c|c|c|c|c|} 
			\hline
			Piece & Files & mean & std & min & max \\
			\hline
			Opus 17, No 4 & 64 & 259.7 & 32.5 & 194.4 & 409.6 \\
			Opus 24, No 2 & 64 & 137.5 & 13.9 & 109.6 & 180.0 \\
			Opus 30, No 2 & 34 & 85.0 & 9.2 & 68.0 & 99.0 \\
			Opus 63, No 3 & 88 & 129.0 & 13.4 & 96.2 & 162.9 \\
			Opus 68, No 3 & 51 & 101.1 & 19.4 & 71.8 & 164.8 \\
			\hline
		\end{tabular}
	\end{center}
\end{table}

We evaluate systems along two axes: alignment accuracy and runtime.  When evaluating the predicted alignment between two recordings A and B, we compare the ground truth beat timestamps in B with the predicted timestamps in B at the corresponding ground truth beat timestamps in A.  Predicted beat timestamps that are within a given error tolerance of the ground truth beat timestamps are considered correct, while predictions outside the allowable error tolerance are considered incorrect.  For a given fixed error tolerance, we can thus calculate an error rate that indicates the percentage of predictions that are incorrect.  By considering a range of error tolerances, we can characterize the tradeoff between error rate and error tolerance.  To evaluate runtime, we construct random feature sequences of length $L$, run the alignment algorithm on the resulting $L \times L$ cost matrix, and average the runtimes across multiple trials.  By considering different values of $L$, we can characterize the runtime in a controlled manner.  All experiments are run on a 2.40 GHz Intel Xeon server with an RTX 3090 GPU.

\section{Results}
\label{sec:results}

We compare the performance of five different alignment algorithms: DTW, FastDTW, and the three Segmental DTW variants.  The basic DTW baseline uses transitions $\{(1,1), (1,2), (2,1)\}$ with multiplicative weights $\{2, 3, 3\}$, which weights all possible alignment paths equally (since all valid alignment paths have the same Manhattan distance).  FastDTW \cite{salvador2007toward} is an approximation of DTW that adopts a multi-resolution approach: it first estimates a global alignment at a low resolution, and then refines the estimated alignment path with successively higher resolutions.  The three Segmental DTW variants -- NSDTW, WSDTW, and SSDTW -- are each evaluated with several different values of $N$ ranging between 2 and 32.  Note that ParDTW produces the exact same alignment as DTW, so its results are not reported separately.

\begin{figure}[!t]
	\centering
	\includegraphics[width=\columnwidth]{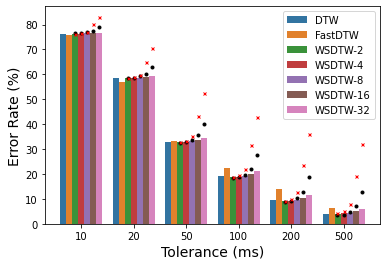}
	\caption{Alignment error rates for DTW, FastDTW, and the 3 variants of Segmental DTW.  From left to right, the bars indicate the performance of DTW, FastDTW, and WSDTW for $N = 2, 4, 8, 16, 32$.  The black dots show results for SSDTW with the same values of $N$.  The red crosses show results for NSDTW with the same values of $N$.}
	\label{fig:results}
\end{figure}

Figure \ref{fig:results} shows the alignment accuracy of all five alignment algorithms. Each group of bars corresponds to a different error tolerance, where we consider error tolerances ranging from 10ms to 500ms.  From left to right, the bars indicate the performance of DTW (blue), FastDTW (orange), and WSDTW with $N=2$ (green), $N=4$ (red), $N=8$ (purple), $N=16$ (brown), and $N=32$ (pink).  The black dots indicate the error rates for SSDTW for the same values of $N$, and the red crosses indicate the error rates for NSDTW for the same values of $N$.

There are 3 things to notice about Figure \ref{fig:results}.  First, WSDTW closely approximates the alignment accuracy of DTW, with a slight degradation in performance as $N$ increases.  For example, with a 50ms error tolerance DTW has an error rate of $32.9\%$, WSDTW with $N=2$ has an error rate of $32.8\%$ and with $N=32$ has an error rate of $34.4\%$.  Second, all three Segmental DTW variants closely approximate DTW for small values of $N$, but the performance of NSDTW and SSDTW degrades rapidly as $N$ increases.  The fact that WSDTW performs best among the three variants is somewhat surprising and unexpected, especially since SSDTW has the strongest guarantees on ordering constraints.  Third, FastDTW has better alignment accuracy than WSDTW with higher values of $N$ at lower error tolerances (and even slightly outperforms regular DTW at very low error tolerances), but has noticeably worse performance than DTW and WSDTW for error tolerances greater than 100ms.

\section{Analyses}
\label{sec:analyses}

In this section we perform four different analyses to better understand the behavior of the proposed alignment algorithms.  These analyses are covered in the next four subsections.

\subsection{Effect of SNR}
\label{subsec:analysis_snr}

The first analysis is to characterize the effect of signal-to-noise ratio (SNR) on alignment accuracy.  While Figure \ref{fig:results} clearly demonstrates that the number of fragments (and thus the length of the fragments) affects alignment accuracy, it is important to recognize that the quality of the approximation also depends on the level of noise or distortion between the two sequences.  For this analysis, we focus on WSDTW since it is the best approximation of DTW among the Segmental DTW variants.

We studied the effect of SNR through a set of controlled experiments.  We first generated several noisy versions of the Mazurka dataset, where noisy versions are constructed by adding additive white Gaussian noise (AWGN) to each audio recording at a fixed SNR.  We consider SNRs of 20 dB, 15 dB, 10 dB, 5 dB, 0 dB, -5 dB, and -10 dB, resulting in 7 noisy versions of the Mazurka dataset.  We then evaluate the accuracy of predicted alignments between clean-noisy pairs of recordings, where each pair consists of one clean recording (i.e.~no noise added) and one noisy recording (i.e.~with AWGN).  Again, we align each (clean) performance to other (noisy) performances of the same mazurka, resulting in the same number of training and testing pairs as before.

\begin{figure}[!t]
	\centering
	\includegraphics[width=\columnwidth]{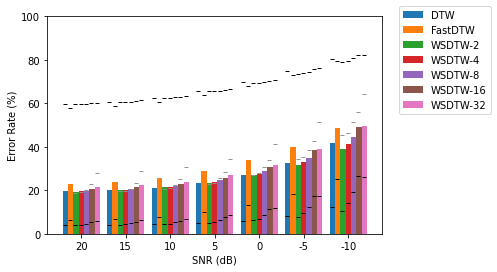}
	\caption{\textcolor{black}{Alignment error rates for aligning clean-noisy pairs, where each pair consists of a clean audio recording and a noisy audio recording with white Gaussian noise added at a fixed SNR. The bars indicate WSDTW error rates with a 100ms error tolerance, and the black horizontal lines indicate WSDTW error rates with 20ms (above) and 500ms (below) error tolerances.  The gray horizontal lines indicate SSDTW error rates with 100ms error tolerance.}}
	\label{fig:analysis_snr}
\end{figure}

Figure \ref{fig:analysis_snr} shows the results of these controlled experiments.  The seven groups of bars correspond to the seven different SNRs ranging from 20 dB to -10 dB.  Within each group, the individual bars indicate the error rates (with 100ms error tolerance) for DTW, FastDTW, and WSDTW with $N=2, 4, 8, 16, 32$.  In addition, black horizontal lines are overlaid on top of each bar to show the error rate for that system with 20ms error tolerance (above) and 500ms error tolerance (below).  \textcolor{black}{The gray horizontal lines indicate the performance of SSDTW with 100ms error tolerance.}

\textcolor{black}{There are three things to notice about Figure \ref{fig:analysis_snr}.  First, we see that WSDTW (colored bars) outperforms SSDTW (gray horizontal lines) across a wide range of SNRs and values of $N$.  Second, across all SNRs the quality of the WSDTW (and SSDTW) approximation becomes worse as $N$ increases.}  This matches the results for our experiments with clean audio.  Third, the WSDTW approximation is worse for lower SNRs.  For example, at 20 dB SNR the 100ms tolerance error rates for DTW ($19.6\%$) and WSDTW ($21.6\%$, $N=32$) are relatively close, whereas at -10 dB the error rates for DTW ($41.7\%$) and WSDTW ($49.8\%$, $N=32$) are quite different.  Thus, there are at least two major factors that affect the quality of the WSDTW approximation: the length of the subsequence fragments and the amount of distortion between the two sequences.  Note that both of these factors contribute to how distinctive the optimal subsequence DTW alignment paths are from suboptimal alignment paths -- longer sequences tend to be more distinctive, and less noise and distortion results in more distinctive alignments.


\begin{table}
	\caption{Comparing alignment runtimes on sequences of varying sizes.  Results are with single-threaded CPU implementations and thus indicate the total amount of computation required.  Times are reported in seconds and are averaged over 10 trials.}
	\label{tab:analysis_runtimes_cpu}
	\begin{center}
		\begin{tabular}{|l|cccccc|} 
			\hline
			& & & \multicolumn{2}{c}{Seq Length} & & \\
			System & 1k & 2k & 5k & 10k & 20k & 50k \\
			\hline
			DTW & 0.016 & 0.086 & 0.57 & 2.36 & 9.55 & 60.6 \\
			ParDTW & 0.029 & 0.11 & 0.85 & 4.06 & 16.9 & 146.8 \\
			FastDTW & 0.69 & 1.38 & 3.45 & 6.90 & 13.8 & 34.6\\
			\hline
			NSDTW-2 & 0.016 & 0.11 & 0.58 & 2.29 & 9.39 & 57.9 \\
			NSDTW-4 & 0.016 & 0.083 & 0.59 & 2.30 & 9.42 & 60.4 \\
			NSDTW-8 & 0.014 & 0.081 & 0.49 & 2.24 & 9.33 & 57.6 \\
			NSDTW-16 & 0.015 & 0.074 & 0.50 & 2.32 & 9.56 & 58.7 \\
			NSDTW-32 & 0.016 & 0.075 & 0.49 & 1.95 & 9.58 & 59.5 \\
			\hline
			WSDTW-2 & 0.016 & 0.085 & 0.57 & 2.38 & 9.42 & 57.0 \\
			WSDTW-4 & 0.014 & 0.077 & 0.57 & 2.28 & 9.49 & 57.7 \\
			WSDTW-8 & 0.015 & 0.082 & 0.50 & 2.38 & 9.61 & 56.5 \\
			WSDTW-16 & 0.015 & 0.074 & 0.47 & 2.15 & 9.58 & 58.9 \\
			WSDTW-32 & 0.017 & 0.076 & 0.49 & 1.90 & 9.63 & 58.0 \\
			\hline
			SSDTW-2 & 0.023 & 0.11 & 0.82 & 3.59 & 20.3 & 136.3 \\
			SSDTW-4 & 0.020 & 0.11 & 0.82 & 3.65 & 15.4 & 147.7 \\
			SSDTW-8 & 0.021 & 0.11 & 0.71 & 3.44 & 14.7 & 103.4 \\
			SSDTW-16 & 0.022 & 0.11 & 0.75 & 3.32 & 13.7 & 93.0 \\
			SSDTW-32 & 0.023 & 0.12 & 0.71 & 2.88 & 13.8 & 91.6 \\
			\hline
		\end{tabular}
	\end{center}
\end{table}

\subsection{Runtime (single-threaded)}
\label{subsec:analysis_runtime_cpu}

\textcolor{black}{The second analysis is to characterize the total amount of computation required for each of the proposed alignment algorithms.  We can accomplish this by measuring runtimes with single-threaded CPU implementations.  This allows us to make a fair comparison of total computation by avoiding confounding factors like the clock speed of the GPU and CPU, the data transfer rate between CPU and GPU, the number of cores on the GPU, and other hardware-dependent characteristics.  In the next subsection (Section \ref{subsec:analysis_runtime_gpu}), we will complement this analysis by comparing wall clock runtimes with parallelized GPU implementations, which are unavoidably affected by the specific details of the hardware setup but are indicative of runtimes that might be expected in practice.}

Table \ref{tab:analysis_runtimes_cpu} compares runtimes of single-threaded CPU implementations of DTW, FastDTW, ParDTW, and all Segmental DTW variants.  Runtimes are measured on pairs of random feature sequences of length 1k, 2k, 5k, 10k, 20k, and 50k, so that the pairwise cost matrix is always an $L \times L$ square matrix.  The three Segmental DTW variants are evaluated with $N = 2, 4, 8, 16, 32$.  Each runtime indicated in the table is the average of 10 trials and is expressed in seconds.

There are a few things to notice about Table \ref{tab:analysis_runtimes_cpu}.  First, we see that DTW, NSDTW, and WSDTW have comparable runtimes across all sequence lengths.  NSDTW adds no additional computation to standard DTW.  WSDTW does add additional computation due to the segment-level operations, but this cost is negligible compared to the amount of computation for frame-level operations.  In fact, in many instances we see that NSDTW and WSDTW actually have lower runtimes than DTW.  Upon further investigation, we found that the primary reason for this is that NSDTW and WSDTW are able to fit matrices into memory cache with higher values of $N$, which results in faster data access.  Second, we see that SSDTW has a significantly higher runtime than the other systems.  This is because of the additional computation required to construct the segment-level transition matrix $T_{seg}$.  Recall that constructing $T_{seg}$ requires backtracing from every possible ending location in every fragment, which results in a significant increase in total amount of computation.  Third, ParDTW has runtimes that are about $1.5$ to $2$ times slower than DTW.  In theory, ParDTW requires the same amount of computation as DTW.  In practice, however, the single-threaded implementation of ParDTW is slower than DTW because it computes the pairwise cost matrix elements individually rather than in a single batch operation (e.g.~a matrix multiplication).

\begin{figure}[!t]
	\centering
	\includegraphics[width=\columnwidth]{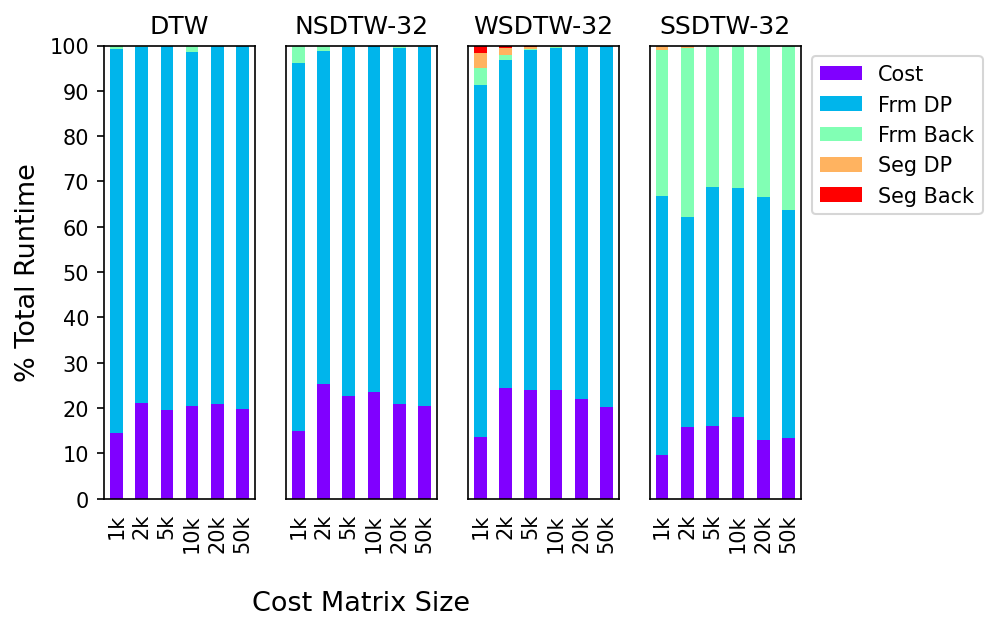}
	\caption{Breakdown of total runtime by component.  The y-axis indicates what percentage of total runtime is due to different stages of computation.  The four panels compare DTW and the three Segmental DTW variants with $N=32$.  Within each panel, the different bars correspond to different sequence lengths.  For the Segmental DTW variants, all stages are parallelizable except for the segment-level dynamic programming and backtracing.}
	\label{fig:analysis_runtime_breakdown}
\end{figure}

Figure \ref{fig:analysis_runtime_breakdown} shows the percentage breakdown of runtime by component for DTW and all Segmental DTW variants.  The three variants of Segmental DTW are evaluated with $N = 32$. The y-axis indicates what percentage of the total runtime is due to five different stages of computation: computing the pairwise cost matrix (``Cost"), frame-level dynamic programming (``Frm DP"), frame-level backtracing (``Frm Back"), segment-level dynamic programming (``Seg DP"), and segment-level backtracing (``Seg Back").  Note that the frame-level operations in Segmental DTW are parallelizable, which corresponds to the purple, blue, and green categories (bottom three components in each bar).  All runtimes are measured on single-threaded CPU implementations.

There are two things to notice about Figure \ref{fig:analysis_runtime_breakdown}.  First, the segment-level operations make up a tiny fraction of the total runtime, especially for longer feature sequences.  Since the segment-level operations are the only ones that are not parallelizable, this indicates that almost all operations in Segmental DTW can be parallelized.  For example, more than $99.66\%$ of the operations in WSDTW-32 are parallelizable for sequences of length 10,000 or longer.  Second, SSDTW expends a much larger fraction of its runtime on frame-level backtracing.  This is because constructing $T_{seg}$ requires backtracing from every possible ending location for every fragment.  The large amount of additional computation that this adds is shown by the green bars in the rightmost panel.

\subsection{Runtime (parallelized)}
\label{subsec:analysis_runtime_gpu}

The third analysis is to consider runtimes with our parallelized GPU implementations of WSDTW and ParDTW.  As mentioned above, it is important to remember that these results are affected by the specific details of the hardware setup, such as the number of cores on the GPU, the clock speed on the GPU, and the data transfer speed between the CPU and GPU.  Nonetheless, these results are useful indicators of the actual wall clock runtime that one might expect in practice.

\begin{table}
	\caption{Effect of $N$ and $M$ on runtime for the GPU-based implementation of WSDTW.  Each number indicates the average runtime in seconds to align two sequences of length 100k.}
	\label{tab:analysis_effectNM}
	\begin{center}
		\begin{tabular}{|l|ccccccc|} 
			\hline
			& & & $M$ & & & &  \\
			$N$ & 1\:\:\:\: & 3\:\:\:\: & 10\:\:\: & 30\:\:\: & 100\:\: & 300\:\: & 1000\: \\
			\hline
			1 & 8.16 & 9.30 & 14.3 & 31.7 & 95.1 & 276 & 915 \\
			3 & 6.79 & 7.25 & 11.5 & 26.0 & 78.0 & 226 & 752 \\
			10 & 6.33 & 5.86 & 7.31 & 13.1 & 33.3 & 92.7 & 301 \\
			30 & 6.26 & 5.27 & 5.51 & 7.25 & 14.3 & 34.9 & 110 \\
			100 & 6.41 & 5.13 & 4.75 & 5.06 & 7.02 & 12.9 & 36.2 \\
			300 & 5.49 & 4.62 & 4.37 & 4.27 & 4.70 & 6.34 & 14.7 \\
			1000 & 5.91 & 5.11 & 4.50 & 4.22 & 3.93 & 4.97 & 9.31 \\
			\hline
		\end{tabular}
	\end{center}
\end{table}

Table \ref{tab:analysis_effectNM} shows the effect of $N$ and $M$ on runtime for the GPU implementation of WSDTW.  Recall that $N$ is the number of fragments to break sequence A into, and $M$ is the number of overlapping chunks to break sequence B into.  The numbers in the table indicate the total runtime required to align two sequences of length 100k with a particular setting of $N$ and $M$.  Each reported time is the average of 10 trials.  Experiments are run with an RTX 3090 with 24 GB of RAM.  Note that different values of $N$ will affect alignment accuracy (as shown in Figure \ref{fig:results}), while different values of $M$ will always yield identical results (but may affect runtime).  \textcolor{black}{Standard deviations for the numbers reported in Table \ref{tab:analysis_effectNM} remained relatively stable, with all entries having standard deviations less than $0.08$s and one entry with $0.11$s standard deviation.}

There are three things to notice about Table \ref{tab:analysis_effectNM}.  First, the runtime is generally faster with a higher value of $N$.  This can be seen by noting that the runtimes in each column generally decrease from top to bottom.  Keep in mind, however, that higher values of $N$ will result in a poorer approximation of DTW, as shown in Figure \ref{fig:results}.  Therefore, the value of $N$ should be selected based on the desired level of approximation, rather than as high as possible.  Second, the runtime only decreases for higher values of $M$ up to some optimal setting.  This can be seen by noting that the runtimes in most rows decrease from left to right, reach a minimum, and then increase.  Higher values of $M$ will result in greater parallelization, but also increase the total amount of computation (since each new chunk must do some redundant computation) and overhead.  For this reason, increasing $M$ beyond the optimal setting will result in longer runtimes.  Since different values of $M$ yield identical predicted alignments, the value of $M$ should be selected to minimize runtime.  Third, the optimal setting for $N$ and $M$ ($N^*=1000$, $M^*=100$) only achieves a modest speedup ($3.93$s) compared to the nominal setting $N=1$, $M=1$ ($8.16$s).  This indicates that most of the benefit from the GPU acceleration comes from parallelizing across diagonals, rather than splitting the pairwise cost matrix into chunks.  This provides a very strong argument for using ParDTW instead of WSDTW, since it guarantees the exact same results as DTW and achieves most of the runtime speedup of WSDTW.

\begin{figure}[!t]
	\centering
	\includegraphics[width=\columnwidth]{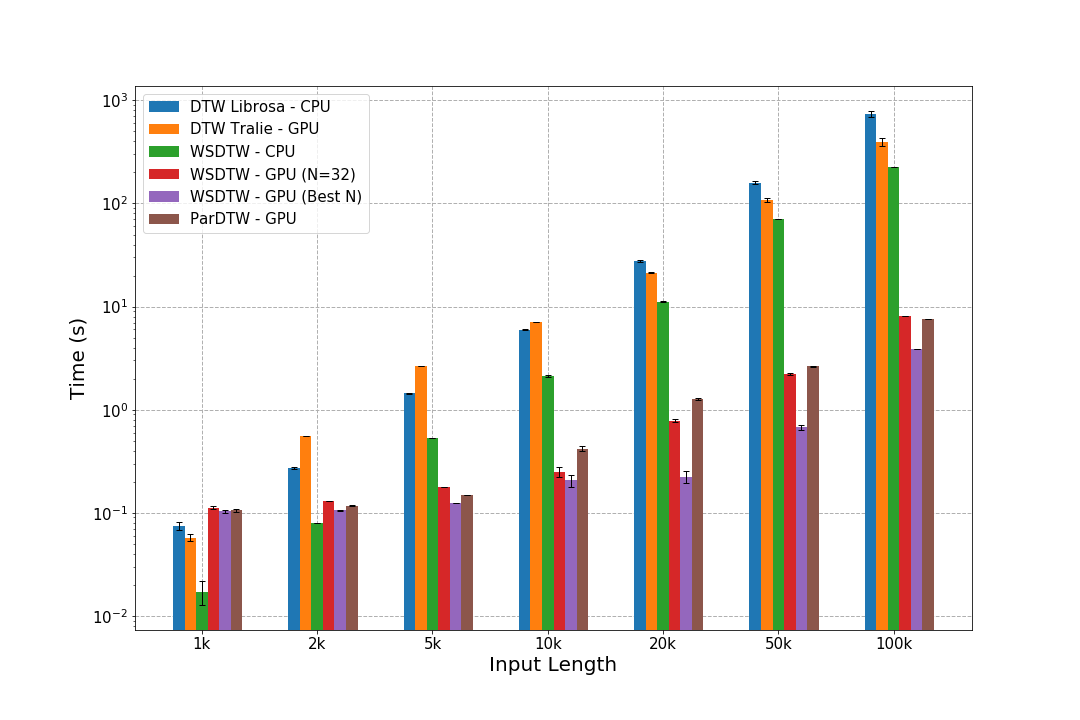}
	\caption{Comparison of runtimes for six different implementations or approximations of DTW: a CPU-based single-threaded implementation of DTW (blue), the previously proposed GPU-based parallelized implementation of exact DTW by Tralie and Dempsey \cite{tralie2020exact} (orange), a CPU-based single-threaded implementation of WSDTW (green), a GPU-based parallelized implementation of WSDTW for $N=32$ (red) as well as for the $N$ that achieves lowest runtime (purple), and a GPU-based parallelized implementation of ParDTW (brown).  \textcolor{black}{The black bars indicate one standard deviation above and below the mean.  Note that some standard deviation bars are so small that they are not visible, and that the log axis results in asymmetric lengths above and below the mean.}}
	\label{fig:analysis_runtime_gpu}
\end{figure}

Figure \ref{fig:analysis_runtime_gpu} compares the runtime of six different implementations or approximations of DTW.  The first implementation is the standard DTW algorithm implemented in numba by the Librosa python library (``DTW Librosa - CPU").  This implementation runs on the CPU in a single thread.  The second implementation is the algorithm proposed by Tralie and Dempsey \cite{tralie2020exact}, which computes exact DTW using a parallelized approach on a GPU (``DTW Tralie - GPU").  The third implementation is an optimized version of WSDTW implemented in cython (``WSDTW - CPU").  This implementation is single-threaded and runs on the CPU.  The fourth and fifth implementations are the GPU versions of WSDTW with $N=32$ (``WSDTW - GPU (N=32)") and with the value of $N$ that achieves the lowest runtime (``WSDTW - GPU (Best N)").  We include $N=32$ as a reasonable setting on the Mazurka dataset, and we include the optimal value of $N$ as a practical lower bound on runtime when alignment accuracy is disregarded.  Both implementations use the optimal setting of $M$ that achieves the lowest runtime.  The sixth implementation is the GPU version of ParDTW (``ParDTW - GPU").  Numbers indicate the average runtime across 10 trials to align two sequences of varying length.  Experiments are run using an RTX 3090 GPU with 24 GB of RAM.

There are three things to notice about Figure \ref{fig:analysis_runtime_gpu}.  First, all of the runtimes increase quadratically with sequence length (which appears as a linear trajectory on a log-log axis).  This is to be expected since all of the implementations have $O(L^2)$ runtime.  Second, the implementation by Tralie and Dempsey only has modest improvements in runtime compared to an optimized, single-threaded CPU implementation of DTW.  Thus, while it addresses the quadratic memory constraints of DTW for long sequences, it does not provide much benefit in terms of runtime.  Third, ParDTW and the GPU version of WSDTW with $N=32$ have similar runtimes, achieving $1.5$ to 2 orders of magnitude speedup for long sequences compared to Librosa.  For example, when aligning sequences of length 100k, the average runtime for Librosa was $572.1$ sec and the average runtime for ParDTW and the GPU implementation of WSDTW with $N=32$ was $6.4$ sec and $7.0$ sec, respectively.

\textcolor{black}{We also investigated the percentage breakdown of wall clock runtime by component.  Across a wide range of sequence lengths (1k to 50k), the percentage breakdown by component was relatively constant for both ParDTW and WSDTW.  For ParDTW, the vast majority of wall clock runtime came from the massively parallelized dynamic programming stage ($99.85\%$) and a tiny percentage came from the non-parallelized backtracing stage ($0.15\%$).  For WSDTW, about $93.9\%$ of the runtime came from the massively parallelized frame-level dynamic programming, $5.9\%$ came from the modestly parallelized frame-level backtracing, and $<0.2\%$ came from the non-parallelized segment-level operations.}

Based on these experimental runtime results, ParDTW is the recommended parallelization strategy.  Compared to WSDTW, ParDTW has several advantages: (a) it is an exact computation of DTW, whereas WSDTW is an approximation, (b) it has no hyperparameters to tune beyond those of regular DTW, whereas WSDTW has two additional hyperparameters that may affect both alignment accuracy and runtime, and (c) for long sequences, it achieves as much speedup as WSDTW.

\subsection{Memory}
\label{subsec:analysis_memory}

The fourth analysis is to characterize the memory requirements of ParDTW.  We focus on the GPU implementation of ParDTW, since it is the recommended parallelization strategy.

We can determine the amount of memory required to align two sequences as follows.  For this discussion, we will assume that the two sequences are of length $L$ with feature dimensionality $D$.  The memory usage on the GPU comes from the following components:
\begin{itemize}
\item Inputs.  Copying the two input sequences to GPU RAM requires $2 \cdot L \cdot D \cdot sizeof(double)$ bytes.
\item Dynamic Programming.  Recall from Section \ref{sec:gpuImpl} that the global pairwise cost matrix is never allocated in memory.  Instead, the pairwise costs are computed on the fly during dynamic programming, and the cumulative cost values are stored in four buffers.  The length of these buffers is the maximum length of a diagonal in the pairwise cost matrix, which for an $L \times L$ matrix is simply $L$.  Therefore, the total memory required for the dynamic programming is approximately $4 \cdot L \cdot sizeof(double)$ bytes.
\item Backtrace Matrix.  The entire backtrace matrix is stored in memory.  This requires $L^2 \cdot sizeof(uint2)$ bytes,\footnote{This is the effective amount of memory required to store the backtrace matrix.  The actual implementation represents the backtrace matrix as an array of uint32 elements and packs 16 consecutive elements into each uint32.} since we only need to store the type of transition at each position.
\item Outputs.  The length of the predicted alignment path is unknown in advance but will be no more than $L$.  Therefore, the predicted alignment (i.e.~sequence of coordinates) will occupy at most $2 \cdot L \cdot sizeof(uint32)$ bytes.
\end{itemize}

Consider memory usage when $L = 250k$ and $D=12$ (e.g.~chroma features).  The inputs take 45 MB ($0.3\%$ of total memory), the dynamic programming takes 7.5 MB ($0.05\%$), the backtrace matrix takes 14.6 GB ($99.6\%$), and the outputs take 2 MB ($0.01\%$).  As this example demonstrates, the backtrace matrix is the dominant factor for memory usage for long sequences.  Therefore, for a GPU with $T$ GB of RAM, the maximum sequence length can be estimated by solving the equation $L_{max}^2 \cdot sizeof(uint2) = T \cdot 2^{30}$, which yields $L_{max} = 2^{16} \cdot \sqrt{T}$.  For a GPU with $T=24$ GB of RAM, this maximum sequence length is approximately $L_{max} =$ 320k.  We have empirically validated that this back-of-the-envelope estimate closely matches the actual maximum sequence length in our GPU implementation.  It is also useful to note that the amount of parallelization ranges between $1$ and $L$ throughout the dynamic programming stage, which results in an average parallelization factor of $\frac{L}{2}$.  This means that the implementation strongly utilizes the GPU's parallelizable architecture when processing long sequences.  

Based on the above memory analysis, we recommend the following metastrategy for selecting an appropriate parallelization strategy for DTW.  For alignment of short sequences, standard single-threaded CPU implementations (e.g.~Librosa) are sufficient.  For longer sequences, one can use a GPU implementation of ParDTW to speed up the runtime by $1.5$ to $2$ orders of magnitude, as long as the computation fits on GPU RAM.  For a GPU with $T$ GB of RAM, this allows for aligning sequences up to length $L_{max} = 2^{16} \cdot \sqrt{T}$.  For sequences longer than $L_{max}$, one can use the Tralie and Dempsey algorithm, which effectively solves the memory problem but is $1.5$ to $2$ orders of magnitude slower than ParDTW.

\section{Conclusion}
\label{sec:concl}

This article explores several parallelizable alternatives to DTW for estimating the alignment between long sequences.  We characterize the performance of these algorithms on an audio-audio alignment task, and we develop GPU-based implementations for the two algorithms with highest alignment accuracy, which we call WSDTW and ParDTW.  Our empirical results indicate that ParDTW is the most practical algorithm for use with GPUs: it computes an exact DTW alignment, reduces runtime by $1.5$ to $2$ orders of magnitude for long sequences compared to current alternatives, and can handle sequence lengths of 250k-320k with typical GPU RAM limits.

\section*{Acknowledgment}
This material is based upon work supported by the National Science Foundation under Grant No. 1948531.

\ifCLASSOPTIONcaptionsoff
  \newpage
\fi



\bibliographystyle{IEEEtran}
\bibliography{SegmentalDTW_taslp}

%
\end{document}